\documentclass[prd,amsfonts,showpacs,nofootinbib]{revtex4}
\usepackage{amsmath}

    \newcommand{\A}{A}
    \newcommand{\C}{C}

    \renewcommand{\H}{\mathcal{H}}
    \renewcommand{\S}{\mathcal{S}}

\begin{document}

\title{Cosmological matching conditions for gravitational waves at second order}
\author{Frederico Arroja$^{1}$\footnote{arrojaf@yukawa.kyoto-u.ac.jp}, Hooshyar Assadullahi$^{2}$\footnote{hooshyar.assadullahi@port.ac.uk}, Kazuya Koyama$^{2}$\footnote{kazuya.koyama@port.ac.uk} and David Wands$^{2}$\footnote{david.wands@port.ac.uk}}
\affiliation{{}$^{1}$Yukawa Institute for Theoretical Physics, Kyoto
University,
Kyoto 606-8502, Japan\\
{}$^{2}$Institute of Cosmology and Gravitation, University of
Portsmouth, Dennis Sciama Building, Burnaby Road, Portsmouth PO1
3FX, United Kingdom}

\begin{abstract}
We compute the second-order matching conditions for tensor metric
perturbations at an abrupt change in the equation of state. For
adiabatic perturbations on large scales the matching hypersurface
coincides with a uniform-density hypersurface. We show that in the
uniform-density gauge both the tensor perturbation and its
time-derivative are continuous in this case. For non-adiabatic
perturbations, the matching hypersurface need not coincide with a
uniform-density hypersurface and the tensor perturbation in the
uniform-density gauge may be discontinuous. However, we show that in
the Poisson gauge both the tensor perturbation and its
time-derivative are continuous for adiabatic or non-adiabatic
perturbations. As an application we solve the evolution equation for
second-order tensor perturbations on large scales for a constant
equation of state and we use the matching conditions to evolve the
solutions through the transition from an inflationary era to a
radiation era. We show that in the radiation era the resulting free
part of the large-scale tensor perturbation (constant mode) is
slow-roll suppressed in both the uniform-density and Poisson
gauges. Thus, we conclude that second-order gravitational waves from
slow-roll inflation are suppressed.
\end{abstract}


\date{\today}
\maketitle

\section{Introduction}

Recent precise measurements of the Cosmic Microwave Background (CMB)
anisotropies have revealed the nature of the primordial
perturbations \cite{Komatsu:2008hk} and future experiments such as
Planck \cite{PLANCK} will improve these measurements significantly.
However, CMB anisotropies and large scale structure probe density
perturbations only on large scales and there is no direct way to
observe density perturbations on smaller scales except for the case
where large density perturbations form primordial black holes
\cite{Green:1997sz, Josan:2009qn}. Recently, it has been recognized
that gravitational waves can be used to probe small scales physics
like preheating \cite{GarciaBellido:2007af, Dufaux:2007pt,
Easther:2007vj} and density perturbations on small scales
\cite{Ananda:2006af,Saito:2008jc}. At first order, tensor
perturbations decouple from density perturbations. However, at
second order, density perturbations can generate tensor
perturbations \cite{Tomita}. The second-order tensor perturbations have been
calculated during inflation \cite{Osano:2006ew}, in the radiation
era \cite{Ananda:2006af} and in the matter era \cite{Tomita,Matarrese:1997ay,Mollerach:2003nq,Baumann:2007zm,
Sarkar:2008ii, Boubekeur:2008kn, Assadullahi:2009nf}. While the
amplitude of the second-order tensor perturbations is generally very
small (determined by the square of the amplitude of density
perturbations), there are several interesting situations where the
second-order tensor perturbations give interesting observational
consequences. For example, if density perturbations are large enough
to form primordial black holes, then these density perturbations can
generate sizable second-order gravitational waves
\cite{Saito:2008jc}. In the matter era, the second-order tensor
perturbation remains constant, which enhances the power spectrum of
the tensor perturbations \cite{Baumann:2007zm}. This happens either
in the late-time matter era or in an early-time matter era due to
the oscillations of the inflaton field after inflation
\cite{Assadullahi:2009nf}. The first-order tensor perturbations
generated during inflation are model dependent and their amplitude
can be very small. However, the amplitude of the second-order tensor
perturbations is determined completely by that of the density
perturbations. Thus this gives a lower bound for the tensor to
scalar ratio \cite{Martineau:2007dj}.

There are still several important issues in the second order
generation of gravitational waves. At second-order, the tensor
perturbations are not gauge invariant under the first-order gauge
transformations \cite{Bruni:1996im}. Thus they depend on the choice
of the gauge for first-order perturbations. It is however possible
to construct gauge-invariant tensor perturbations by eliminating the
gauge degrees of freedom
\cite{Mukhanov:1996ak,Nakamura:2003wk,Malik:2003mv,Malik:2008yp,
Malik:2008im}. A commonly used gauge is the Poisson gauge
\cite{Bruni:1996im,Matarrese:1997ay,Nakamura:2006rk} and almost all calculations of second-order
gravitational waves have been done in the Poisson gauge. During
inflation, another commonly used gauge is the uniform-density gauge.
This gauge is often used to calculate higher-order correlation
functions for the curvature perturbation \cite{Maldacena:2002vr} and
these calculations involve the second-order tensor perturbations
because they are coupled to the curvature perturbations at higher
order \cite{Arroja:2008ga}. In fact, the amplitudes of tensor
perturbations in these two gauges can be very different. In the
Poisson gauge, the amplitude of the second-order tensor
perturbations during inflation is suppressed by slow-roll parameters
but it is not in the uniform density gauge.

The question is then whether the amplitude of gravitational waves
{\em after} inflation is slow-roll suppressed, or indeed whether
large tensor perturbations can be generated at second order by the
transition to a radiation dominated universe. In order to address
this problem, we should track the evolution of the tensor
perturbations during the transition from inflation to the radiation
era.

At first-order, matching conditions for cosmological perturbations
have been developed in cases where there is an abrupt change in the
expansion rate at the matching surface
\cite{Hwang:1991an,Deruelle:1995kd,Martin:1997zd, Copeland:2006tn},
such as reheating at the end of inflation. In this paper, we extend
this matching condition to second-order perturbations and derive
matching conditions for second-order tensor perturbations on large
scales. Matching conditions will be derived in two cases. For
adiabatic perturbations on large scales a sudden transition in the
equation of state must occur at a specific density. Thus we apply
the matching conditions across a uniform-density hypersurface. We
will also consider the case where the transition surface is given by
a fixed value of some field $\chi$, which need no longer coincide
with a uniform-density hypersurface in the presence of non-adiabatic
perturbations. In this case the matching conditions are applied
across a uniform-$\chi$ hypersurface. These matching conditions are
applied to the second-order tensor perturbations generated during
inflation and we calculate the second-order tensor perturbations in
the radiation era after reheating in both the Poisson gauge and the
uniform-density gauge. We will show that the constant mode of the
second-order gravitational waves on large scales, corresponding to
the free part of the tensor perturbations, is the same in the two
gauges in the radiation era and it is suppressed by slow-roll
parameters.

This paper is organized as follows. In section II we summarize the
gauge transformations at the first and second-order and derive
solutions for the metric perturbations on large scales in the
Poisson gauge and in the uniform-density gauge. In section III, the
evolution equation for the second-order tensor perturbations is
derived. General solutions for the second-order tensor perturbations
are derived on large scales with a constant equation of state and we
check the consistency of solutions in two different gauges under the
gauge transformation. In section IV, matching conditions for
perturbations are developed at the first and second-order in the
cases where there is an abrupt change in the expansion rate at the
matching surface. We consider both adiabatic and non-adiabatic
matchings. In section V, we apply the matching conditions to the
second-order tensor perturbations generated during slow-roll
inflation and calculate the evolution of the second-order tensor
perturbations on large scales in the radiation era. Section V is
devoted to conclusions.

\section{Gauge-invariant perturbations}

In this section, after introducing the necessary notation and the
perturbed metric, we shall review the first and second-order general
gauge transformations for the perturbations. After that, using some
simplifying assumptions, we will obtain the gauge transformation
rules for second-order tensor perturbations. We will also derive the
solutions for first-order scalar metric perturbations on large
scales.

\subsection{Perturbed metric}

Throughout this work Greek indices $\mu,\nu,\ldots$ take values
from 0 to 3, while Latin indices $i,j,\ldots$ denote spatial
indices and can take values from 1 to 3. The components of a
perturbed spatially flat Friedman-Robertson-Walker metric (FRW) can be written
as
\begin{eqnarray}
g_{00}&=&-a(\eta)^2\left(1+2\sum_{r=1}^{+\infty}
\frac{1}{r!}A^{(r)}\right),
\\
g_{0i}&=&a(\eta)^2\sum_{r=1}^{+\infty} \frac{1}{r!}B_i^{(r)},
\\
g_{ij}&=&a(\eta)^2\left[\left(1-2\sum_{r=1}^{+\infty}
\frac{1}{r!}C^{(r)}\right)\delta_{ij}+\sum_{r=1}^{+\infty}
\frac{1}{r!}C^{(r)}_{ij}\right],
\end{eqnarray}
where $\eta$ denotes conformal time and $C^{(r)}_{ij}$ is
traceless, i.e. $C^{i(r)}_i=0$, where the index was raised with
$\delta^{ij}$. The superscript $(r)$ indicates the order of the
perturbation. The perturbations depend on conformal time and
position $\mathbf{x}$. The inverse metric up to second order can be found in Appendix. We decompose the perturbations in scalar,
vector and tensor parts as
\begin{equation}
B_i^{(r)}=\partial_iB^{(r)}+S^{(r)}_i,
\end{equation}
where $\partial^iS^{(r)}_i=0$.
\begin{equation}
C_{ij}^{(r)}=2D_{ij}E^{(r)}+\partial_iF_j^{(r)}+\partial_jF_i^{(r)}+h_{ij}^{(r)},
\end{equation}
where $\partial^iF^{(r)}_i=0$ and
$h
^{i(r)}_i=\partial^ih^{(r)}_{ij}=0$. The operator
$D_{ij}$ is defined as
$D_{ij}=\partial_i\partial_j-\frac{1}{3}\delta_{ij}\partial^2$.
The energy density $\rho$ and the four velocity of matter are
perturbed like
\begin{equation}
\rho=\rho_{(0)}+\sum_{r=1}^{+\infty} \frac{1}{r!}\delta\rho^{(r)},
\end{equation}
\begin{equation}
u^\mu=\frac{1}{a}\left(\delta^\mu_0+\sum_{r=1}^{+\infty}
\frac{1}{r!}v^{\mu(r)}\right).
\end{equation}
$u^\mu$ is normalized as $u^\mu u_\mu=-1$ and this implies that
$v^{0(r)}$ is related to the lapse perturbation $\A^{(r)}$. The
velocity perturbation is decomposed as
\begin{equation}
v^{i(r)}=\partial^iv^{\|(r)}+v^{\bot i(r)}
\end{equation}

\subsection{Gauge transformations}

Bruni \emph{et al.} \cite{Bruni:1996im} have shown us that given a tensor $T$, its
perturbations in two different gauges $X$, and $Y$ are related at
first order as
\begin{equation}
{\delta T^Y}^{(1)}-{\delta T^X}^{(1)}=\pounds_{\xi}^{(1)}T^{(0)},\label{GT1}
\end{equation}
and at second order we have
\begin{equation}
{\delta T^Y}^{(2)}-{\delta
T^X}^{(2)}=\left(\pounds_{\xi^{(2)}}+\pounds^2_{\xi^{(1)}}\right)T^{(0)}+2\pounds_{\xi^{(1)}}{\delta
T^X}^{(1)},
\end{equation}
where $\xi^{(1)}$ and $\xi^{(2)}$ are the gauge shifts at first
and second order respectively. The Lie derivative along a vector
field $\xi$ of a rank two covariant tensor $T$ is given by
$\pounds_\xi
T_{\mu\nu}=T_{\mu\nu,\lambda}\xi^\lambda+\xi^\lambda_{,\mu}T_{\lambda\nu}+\xi^\lambda_{,\nu}T_{\mu\lambda}$.
We also decompose $\xi^{\mu(r)}$ is scalar and vector parts as
\begin{equation}
\xi^{0(r)}=\alpha^{(r)}, \quad
\xi^{i(r)}=\partial^i\beta^{(r)}+d^{i(r)},
\end{equation}
where $\partial_id^{i(r)}=0$.

\subsubsection{First order}

At first order, the gauge transformations, Eq. (\ref{GT1}), can be written as
\begin{eqnarray}
\tilde\A^{(1)}&=&\A^{(1)}+{\alpha^{(1)}}'+\mathcal{H}\alpha^{(1)},\label{GTs}
\\
\tilde\C^{(1)}&=&\C^{(1)}-\frac{1}{3}\partial^2\beta^{(1)}-\mathcal{H}\alpha^{(1)},
\\
 \tilde
B^{(1)}_i&=&B^{(1)}_i-\alpha_{,i}^{(1)}+{\beta_{,i}^{(1)}}'+{d_i^{(1)}}',
\\
\tilde
C_{ij}^{(1)}&=&C_{ij}^{(1)}+2D_{ij}\beta^{(1)}+d_{i,j}^{(1)}+d_{j,i}^{(1)},\label{GTe}
\end{eqnarray}
where prime denotes the derivative with respect to conformal time
and the conformal Hubble rate is $\mathcal{H}=\frac{a'}{a}$.
The last two equations imply
\begin{eqnarray}
 \label{tildeB}
\tilde B^{(1)}&=&B^{(1)}-\alpha^{(1)}+{\beta^{(1)}}',
\\
\tilde S^{(1)}_i&=&S^{(1)}_i+{d_i^{(1)}}',
\\
 \label{tildeE}
\tilde E^{(1)}&=&E^{(1)}+\beta^{(1)},
\\
\tilde F^{(1)}_i&=&F^{(1)}_i+d_i^{(1)},
\\
\tilde h_{ij}^{(1)}&=&h_{ij}^{(1)}.
\end{eqnarray}
Four dimensional scalars, like the energy density or the pressure,
have the transformation rule
\begin{equation}
 \label{tilderho}
\delta\tilde{\rho}=\delta\rho+{\rho^{(0)}}'\alpha^{(1)}, \quad
\delta\tilde{P}=\delta P+{P^{(0)}}'\alpha^{(1)}.
\end{equation}
The spatial component of the four velocity transform as
\begin{equation}
\tilde v^{i(1)}=v^{i(1)}-\partial^i{\beta^{(1)}}'-{d^{i(1)}}',
\end{equation}
or equivalently
\begin{equation}
\tilde v^{\|(1)}=v^{\|(1)}-{\beta^{(1)}}', \quad \tilde v^{\bot
i(1)}=v^{\bot i(1)}-{d^{i(1)}}',
\end{equation}
while the temporal component transforms as $\A^{(1)}$ because of
the normalization constraint $v^{0(1)}=-\A^{(1)}$.

Following Bardeen \cite{Bardeen:1980kt}, we can define gauge
invariant quantities by specifying completely the choice of
coordinates. The longitudinal or Poisson gauge is defined at first
order by requiring $B_P^{(1)}=E_P^{(1)}=0$, which from
Eqs.~(\ref{tildeB}) and~(\ref{tildeE}) fixes
$\beta_P^{(1)}=-E^{(1)}$ and $\alpha_P^{(1)}=B^{(1)}-E^{(1)\prime}$,
where the subscript $P$ denotes the Poisson gauge quantity. The
remaining scalar metric perturbations then have the gauge invariant
definitions
\begin{eqnarray}
\Phi&\equiv& A_P^{(1)} = A^{(1)}+\mathcal{H}\left(B^{(1)}-E^{(1)'}\right)+\left(B^{(1)}-E^{(1)'}\right)',\\
\Psi&\equiv& C_P^{(1)} =
C^{(1)}-\mathcal{H}\left(B^{(1)}-E^{(1)'}\right)+\frac{1}{3}\partial^2E^{(1)},
\end{eqnarray}
The gauge-invariant density perturbation corresponding to density
perturbations in this specific gauge is, from~Eq.(\ref{tilderho}),
\begin{equation}
\delta\rho_P^{(1)} = \delta\rho +
{\rho^{(0)\prime}}\left(B^{(1)}-E^{(1)'}\right) \,.
\end{equation}

An alternative set of gauge invariant variables can be constructed
by working on uniform-density hypersurfaces for which
$\delta\rho_{UD}=0$, which from Eq.~(\ref{tilderho}) fixes
$\alpha_{UD}^{(1)}=-\delta\rho/\rho^{(0)\prime}$.
The curvature perturbation in the uniform density gauge is thus the
gauge invariant variable
\begin{equation}
-\zeta\equiv \C^{(1)}_{UD} + \frac{1}{3}\partial^2E_{UD}^{(1)} =
\C^{(1)} + \frac{1}{3}\partial^2E^{(1)} +
\frac{\mathcal{H}}{{\rho^{(0)}}'}\delta\rho^{(1)}, \label{zetadef}
\end{equation}
Here and in the rest of this work we have choose the threading of
the time slices, $\beta_{UD}^{(1)}=-E^{(1)}$, such that
$E^{(1)}_{UD}=0$. The gauge invariant definition of the scalar part
of the shift vector (\ref{tildeB}) in the uniform density gauge is
given as
\begin{equation}
B_{UD}^{(1)}= B^{(1)} - E^{(1)\prime} +
\frac{\delta\rho^{(1)}}{\rho^{(0)\prime}}
 =
\frac{\delta\rho_P^{(1)}}{\rho^{(0)\prime}}
 =
-\left( \frac{\zeta+\Psi}{\mathcal{H}} \right).
 \label{scalarshiftUD}
\end{equation}

In the absence of anisotropic stress, the Einstein equations in the
Poisson gauge require $\Phi=\Psi$ and the energy constraint equation
gives a relation between the curvature perturbation in the
uniform-density and Poisson gauges
\begin{equation}
\zeta=-\frac{2\mathcal{H}^2-\mathcal{H}'}{\mathcal{H}^2-\mathcal{H}'}\Psi-\frac{\mathcal{H}}{\mathcal{H}^2-\mathcal{H}'}\Psi'+\frac{1}{3\left(\mathcal{H}^2-\mathcal{H}'\right)}\partial^2\Psi.
 \label{zetaeq}
\end{equation}
In the next section, when writing the equations of motion for second
order tensor perturbation, we will neglect terms with more than two
spatial gradients. We also ignore non-adiabatic perturbations. This
implies that we can take $\zeta'=0$ at zeroth order in
the gradient expansion. The scalar part of the Einstein equations gives
\begin{equation}
a^{-2} \left( 6 {\cal H} (C^{(1)'}+ {\cal H} A^{(1)}) -2 \partial^2 C^{(1)} \right)
=-\kappa_G^2 \delta \rho^{(1)},
\label{scalarEinstein}
\end{equation}
where $\kappa_G$ is the gravitational constant. In the uniform-density gauge $\delta \rho^{(1)}=0$,
thus on large scales $A_{UD}=0$.
For a perfect fluid with a constant equation of state $w$ we have
$\zeta=V=$constant on large scales, and Eq.~(\ref{zetaeq}) gives the
solution
\begin{equation}
\Psi=-\frac{3+3w}{5+3w}V+W\eta^{-\frac{5+3w}{1+3w}}, \label{Phisol}
\end{equation}
where $W=W(\mathbf{x})$ is a constant of integration to be determined by the
first order junction conditions. If $w=1/3$, the previous equation
simplifies to give
\begin{equation}
\Psi=-\frac{2}{3}V+W\eta^{-3}.
\end{equation}

\subsubsection{Second order}

The second-order gauge transformation can be found in Bruni \emph{et al.} For the trace-free spatial metric perturbation it reads
\begin{eqnarray}
\tilde{C}^{(2)}_{ij} & = &C^{(2)}_{ij}
+2\left({C^{(1)}_{ij}}'
+2\mathcal{H}C^{(1)}_{ij}\right)\alpha^{(1)}
+2C^{(1)}_{ij,k}\xi^{k(1)}
\nonumber \\
&&+ 2\left(-4\C^{(1)}+\alpha^{(1)}\partial_0
+\xi^{k(1)}\partial_k+4\mathcal{H}\alpha^{(1)}\right)
\left(d^{(1)}_{(i,j)}+D_{ij}\beta^{(1)}\right)
\nonumber \\
&&+ 2\left[\left(2B^{(1)}_{(i}-\alpha^{(1)}_{,(i}
+{\xi^{(1)}_{(i}}'\right) \alpha^{(1)}_{,j)}
-\frac{1}{3}\delta_{ij} \left(2B^{k(1)}-\alpha^{,k(1)}
+{\xi^{k(1)}}'\right)\alpha^{(1)}_{,k}\right]
\nonumber\\
&&+ 2\left[\left(2C^{(1)}_{(i|k|}
+\xi^{(1)}_{k,(i}+\xi^{(1)}_{(i,|k|}\right) \xi^{(1)k}_{,j)}
-\frac{1}{3}\delta_{ij}\left(2C^{(1)}_{lk}
+\xi^{(1)}_{k,l}+\xi^{(1)}_{l,k}\right) \xi^{k,l(1)}\right]
\nonumber \\
&& +2\left(d^{(2)}_{(i,j)} +D_{ij}\beta^{(2)}\right).
\label{GT3dtraceless}
\end{eqnarray}
We can define gauge invariant second order tensor perturbations in the Poisson gauge as
\begin{eqnarray}
 \label{h2defP}
{\tilde h}{^{(2)}_{ij}}_P & = &h^{(2)}_{ij}
+2\bigg[\left({C^{(1)}_{ij}}'
+2\mathcal{H}C^{(1)}_{ij}\right)\alpha^{(1)}
+C^{(1)}_{ij,k}\xi^{k(1)}
\nonumber \\
&&+ \left(-4\C^{(1)}+\alpha^{(1)}\partial_0
+\xi^{k(1)}\partial_k+4\mathcal{H}\alpha^{(1)}\right)
\left(d^{(1)}_{(i,j)}+D_{ij}\beta^{(1)}\right)
\nonumber \\
&&+ \left(2B^{(1)}_{(i}-\alpha^{(1)}_{,(i}
+{\xi^{(1)}_{(i}}'\right) \alpha^{(1)}_{,j)}
+ \left(2C^{(1)}_{(i|k|}
+\xi^{(1)}_{k,(i}+\xi^{(1)}_{(i,|k|}\right) \xi^{(1)k}_{,j)}
\bigg]^{TT},
\end{eqnarray}
where all the quantities in the r.h.s. of the previous equation are in an arbitrary gauge and the parameters of the gauge transformations are explicitly given as $\alpha^{(1)}=B^{(1)}-E^{(1)'}$, $\beta^{(1)}=-E^{(1)}$, $d_i^{(1)}=-F_i^{(1)}$ and $\xi^{i(1)}=\partial^i\beta^{(1)}+d^{i(1)}$.
In a similar way, a gauge invariant definition of the uniform density second order tensor perturbation is
\begin{eqnarray}
 \label{h2defUD}
{\tilde h}{^{(2)}_{ij}}_{UD} & = &h^{(2)}_{ij}
+2\bigg[\left({C^{(1)}_{ij}}'
+2\mathcal{H}C^{(1)}_{ij}\right)\alpha^{(1)}
+C^{(1)}_{ij,k}\xi^{k(1)}
\nonumber \\
&&+ \left(-4\C^{(1)}+\alpha^{(1)}\partial_0
+\xi^{k(1)}\partial_k+4\mathcal{H}\alpha^{(1)}\right)
\left(d^{(1)}_{(i,j)}+D_{ij}\beta^{(1)}\right)
\nonumber \\
&&+ \left(2B^{(1)}_{(i}-\alpha^{(1)}_{,(i}
+{\xi^{(1)}_{(i}}'\right) \alpha^{(1)}_{,j)}
+ \left(2C^{(1)}_{(i|k|}
+\xi^{(1)}_{k,(i}+\xi^{(1)}_{(i,|k|}\right) \xi^{(1)k}_{,j)}
\bigg]^{TT},
\end{eqnarray}
where $\alpha^{(1)}=-\delta\rho^{(1)}/{\rho^{(0)}}'$, $\beta^{(1)}=-E^{(1)}$, $d_i^{(1)}=-F_i^{(1)}$ and $\xi^{i(1)}=\partial^i\beta^{(1)}+d^{i(1)}$.

In this work, we will neglect the effect of first-order vector and tensor perturbations in order to simplify the second-order equations as much as possible. In many models of the very early universe, such as slow-roll inflation which we consider later, vector and tensor perturbations are suppressed with respect to scalar perturbations. First-order primordial vector and tensor perturbations, if they do exist, would provide an additional, independent source of second-order gravitational waves.

We choose to set $E^{(r)}$ to zero in both Poisson and uniform density gauges. This implies that in the gauge transformations between these two gauges the parameter $\beta^{(1)}$ is zero (and $d^{i(1)}=0$).
The first order temporal gauge shift, $\alpha^{(1)}$, is given in
Table I. For example, if we start in the Poisson gauge and if we do a
gauge change with $\alpha^{(1)}=-\delta\rho^{(1)}/{\rho^{(0)}}'$ then we obtain the quantity in the uniform density gauge (tilde gauge), as
given in Eqs. (\ref{GTs}-\ref{GTe}).
\begin{table}
\caption{Time gauge shift $\alpha^{(1)}$ at first order between
our different gauges.}
\begin{tabular}{c|c|c}
$\alpha^{(1)}$ &\hspace{.5cm}Poisson\hspace{.5cm}
&\hspace{.2cm}Uniform
density\hspace{.2cm}
\\\hline
\rule[-3mm]{0mm}{8mm}
Poisson &0 &$-\delta\rho^{(1)}/{\rho^{(0)}}'$    \\
\rule[-3mm]{0mm}{8mm}
Uniform density &$B^{(1)}$   &0
\end{tabular}
\end{table}
The Poisson gauge is defined by $B^{(1)}=E^{(1)}=F_i^{(1)}=0$ and the
uniform density gauge is $\delta\rho^{(1)}=E^{(1)}=F_i^{(1)}=0$.
With the previous assumptions and with the choice $E^{(1)}=0$, the gauge transformation equation (\ref{GT3dtraceless}) for the
traceless perturbation $C_{ij}^{(2)}$ simplifies considerably to
\begin{eqnarray}
\tilde{C}^{(2)}_{ij} & = &C^{(2)}_{ij} +
2\left[\left(2B^{(1)}_{(i}-\alpha^{(1)}_{,(i} \right)
\alpha^{(1)}_{,j)} -\frac{1}{3}\delta_{ij}
\left(2B^{k(1)}-\alpha^{,k(1)}\right)\alpha^{(1)}_{,k}\right]
+2\left(d^{(2)}_{(i,j)} +D_{ij}\beta^{(2)}\right),
\end{eqnarray}
which gives the following rule for transforming second order tensors
\begin{eqnarray}
\tilde{h}^{(2)}_{ij} & = &h^{(2)}_{ij} +
2\left[\left(2B^{(1)}_{(i}-\alpha^{(1)}_{,(i} \right)
\alpha^{(1)}_{,j)}\right]^{TT},\label{GT}
\end{eqnarray}
where $TT$ means the transverse and traceless part of the terms in
between square brackets. In the case of transforming from the Poisson gauge to the uniform density gauge the previous equation further simplifies to
give
\begin{eqnarray}
\tilde{h}{^{(2)}_{ij}}_{UD}
 - {h}{^{(2)}_{ij}}_P & = &
\frac{2}{\mathcal{H}^2}
\left[(\zeta+\Psi)(\zeta+\Psi)_{,ij}\right]^{TT}. \label{GTPUD}
\end{eqnarray}
The right-hand-side of this equation leads to different solutions for second-order tensor perturbation in uniform-density or Poisson gauges. Nonetheless we have constructed gauge-invariant expressions, (\ref{h2defP}) and~(\ref{h2defUD}), for the second-order tensor perturbation in the uniform-density and Poisson gauges, and the difference between them is also manifestly gauge-invariant.


\section{Tensor perturbation equations}

In this section, we will solve the first and second-order large-scale equations of motion for tensor perturbations in a fluid with a constant equation of state, in both the Poisson and the uniform-density gauges. At the end of the section, we shall present the second-order gauge transformations and the relations between the integration constants that appear in the two gauges.

\subsection{First order}

The first order tensor perturbation is gauge invariant and its equation of motion on large scales is:
\begin{equation}
{h_{ij}^{(1)}}''+2\mathcal{H}{h_{ij}^{(1)}}'=0,
\end{equation}
that can be easily solved
\begin{equation}
h_{ij}^{(1)}=Y_{ij}+Z_{ij}\eta^{-\frac{3-3w}{1+3w}},
\end{equation}
where $Y_{ij}$ and $Z_{ij}$ are transverse and traceless integration constants to be determined by the initial conditions.
During matter or radiation eras, $Z_{ij}$ is a decaying mode and the growing mode is constant in time, as expected for first-order gauge invariant tensor perturbations on super-horizon scales.

\subsection{Second order}

In this subsection we will present the equation of motion for second
order tensor perturbations in the uniform density gauge and in the
Poisson gauge.
Note that we neglect first order vector and tensor perturbations
and we choose $E^{(r)}=0$. We also ignore first
and second order anisotropic stress.

Using the second order Einstein equations from Bartolo
\emph{et al.} review \cite{Bartolo:2004if} we get
\begin{equation}
{{G}_j^{i(2)}}^{TT}=\kappa_G^2{T_j^{i(2)}}^{TT},\label{Einsteineq}
\end{equation}
where
\begin{equation}
{T_j^{i(2)}}^{TT}=\left(\rho^{(0)}+P^{(0)}\right)\left[\partial^iv^{\|(1)}\left(\partial_jv^{\|(1)}+\partial_jB^{(1)}\right)\right]^{TT},
\end{equation}
\begin{eqnarray}
{{G}_j^{i(2)}}^{TT}&=&a^{-2}\left(\frac{1}{4}{h_j^{i(2)}}''+\frac{1}{2}\mathcal{H}{h_j^{i(2)}}'-\frac{1}{4}\partial^2h_j^{i(2)}\right)
\nonumber
\\&&+a^{-2}\Big[\partial^i\A\partial_j\A+2\A\partial^i\partial_j\A-2\C\partial^i\partial_j\A-\partial_j\A\partial^i\C-\partial^i\A\partial_j\C+3\partial^i\C\partial_j\C+4\C\partial^i\partial_j\C
\nonumber
\\&&+2\mathcal{H}\partial^i B\partial_j\A+4\mathcal{H}\A\partial^i\partial_j B+\A'\partial^i\partial_j B+2\A\partial^i\partial_j B'
+\partial^2 B\partial^i\partial_jB-\partial_j\partial^kB\partial^i\partial_kB
\nonumber \\&&-2\mathcal{H}\partial^i\C\partial_jB
-2\mathcal{H}\partial^iB\partial_j\C-\partial^i\C'\partial_jB+\partial_j\C'\partial^iB
-\partial^i\C\partial_j B'-\partial_j\C\partial^i B'-2\C\partial^i\partial_j B'
\nonumber
\\&&+\C'\partial^i\partial_jB-4\mathcal{H}\C\partial_j\partial^iB
\Big]^{TT},
\end{eqnarray}
where in the previous equation we omitted the superscript $(1)$ to
indicate the order of the perturbations and we corrected the typos of equation (A.43) of Ref. \cite{Bartolo:2004if}.

\subsubsection{Poisson gauge}

The equation of motion for second-order tensor perturbations in the
Poisson gauge is known and can be found in Baumann \emph{et al.}
\cite{Baumann:2007zm} (see also \cite{Ananda:2006af,Nakamura:2006rk,Mangilli:2008bw}). Under our
assumptions (for constant $w$) it simplifies to
\begin{equation}
{h_{ij}^{(2)}}''_P+2\mathcal{H}{h_{ij}^{(2)}}'_P-\partial^2{h_{ij}^{(2)}}_P= 8\left[\partial_i\Psi\partial_j\Psi+\frac{2}{3\left(1+w\right)\mathcal{H}^2} \partial_i\left(\Psi'+\mathcal{H}\Psi\right)\partial_j\left(\Psi'+\mathcal{H}\Psi\right)\right]^{TT}.
\label{EomP1}
\end{equation}
Using the solution for $\Psi$ as in equation (\ref{Phisol}), the equation of motion can be
written in a simple form as
\begin{equation}
{h_{ij}^{(2)}}''_P+2\mathcal{H}{h_{ij}^{(2)}}'_P = 8\frac{3+3w}{5+3w}\left[\partial_iV\partial_jV\right]^{TT}+4(5+3w)\left[\partial_iW\partial_jW\right]^{TT}\eta^{-2\frac{5+3w}{1+3w}}.
\label{eomP2}
\end{equation}
where we drop the gradient terms on the left-hand-side of (\ref{EomP1}). This is the equation of motion for second-order tensor perturbations, at leading order in a gradient expansion.
It can be solved to give
\begin{equation}
{h_{ij}^{(2)}}_P={Y_{ij}}_P+{Z_{ij}}_P\eta^{-\frac{3-3w}{1+3w}}
+4\frac{(1+3w)(3+3w)}{(5+3w)^2}\left[\partial_iV\partial_jV\right]^{TT}\eta^2 +\frac{(1+3w)^2}{2}\left[\partial_iW\partial_jW\right]^{TT}\eta^{-\frac{8}{1+3w}},
\label{solP}
\end{equation}
where ${Y_{ij}}_P$ and ${Z_{ij}}_P$ are integration constants.

\subsubsection{Uniform density gauge}

Using Einstein equations (\ref{Einsteineq}), we can derive the equation of motion for second-order tensor perturbations in the uniform-density gauge as

\begin{equation}
{h_j^{i(2)}}''_{UD}+2\mathcal{H}{h_j^{i(2)}}'_{UD}-\partial^2{h_j^{i(2)}}_{UD}=-4S_{ij}^{TT},\label{EinsteineqUD}
\end{equation}
where the source is given by
\begin{eqnarray}
S_{ij}&=&\partial^i\A\partial_j\A+2\A\partial^i\partial_j\A+2\zeta\partial^i\partial_j\A+\partial_j\A\partial^i\zeta+\partial^i\A\partial_j\zeta+3\partial^i\zeta\partial_j\zeta+4\zeta\partial^i\partial_j\zeta
+2\mathcal{H}\partial^i B\partial_j\A+4\mathcal{H}\A\partial^i\partial_j B
\nonumber
\\&&+\A'\partial^i\partial_j B+2\A\partial^i\partial_j B'
+\partial^2 B\partial^i\partial_jB-\partial_j\partial^kB\partial^i\partial_kB
+2\mathcal{H}\partial^i\zeta\partial_jB
+2\mathcal{H}\partial^iB\partial_j\zeta+\partial^i\zeta'\partial_jB-\partial_j\zeta'\partial^iB
\nonumber \\&&
+\partial^i\zeta\partial_j B'+\partial_j\zeta\partial^i B'+2\zeta\partial^i\partial_j B'
-\zeta'\partial^i\partial_jB+4\mathcal{H}\zeta\partial_j\partial^iB
-\frac{3}{2}\partial_j\left(-\zeta'+\mathcal{H}A\right)\partial^{-2}\partial^i\left(\zeta'+\frac{\mathcal{H}}{\rho^{(0)}+P^{(0)}}\delta P\right),\nonumber\\
\end{eqnarray}
and we have used the first-order Einstein equations to simplify the result.
Using Eqs.~(\ref{scalarshiftUD}), (\ref{Phisol}), $A_{UD}=0$ on large scales, ignoring higher derivatives terms and the gradient term
on the left-hand-side of (\ref{EinsteineqUD}),
we can write the equation of motion (\ref{EinsteineqUD}) for
second-order tensor perturbations in the uniform-density gauge
at leading order in a gradient expansion
as
\begin{equation}
{h_{ij}^{(2)}}''_{UD}+2\mathcal{H}{h_{ij}^{(2)}}'_{UD}=4\left[\partial_iV\partial_jV\right]^{TT}.
\label{eomUD}
\end{equation}
This can be easily solved
\begin{equation}
 \label{hUD}
{h_{ij}^{(2)}}_{UD}={Y_{ij}}_{UD}+{Z_{ij}}_{UD}\eta^{-\frac{3-3w}{1+3w}}+2\frac{1+3w}{5+3w}\left[\partial_iV\partial_jV\right]^{TT}\eta^2,
\end{equation}
where ${Y_{ij}}_{UD}$ and ${Z_{ij}}_{UD}$ are integration constants to be determined by the initial conditions.

The second-order gauge transformation (\ref{GTPUD}) between these two
gauges is
\begin{eqnarray}
{h_{ij}^{(2)}}_{UD}={h_{ij}^{(2)}}_{P}&&\!\!\!\!
-2\left(\frac{1+3w}{5+3w}\right)^2\left[\partial_iV\partial_jV\right]^{TT}\eta^2 -\frac{(1+3w)^2}{2}\left[\partial_iW\partial_jW\right]^{TT}\eta^{-\frac{8}{1+3w}}
\nonumber\\&&-\frac{(1+3w)^2}{5+3w}\left[\partial_iW\partial_jV+\partial_jW\partial_iV\right]^{TT}\eta^{\frac{-3+3w}{1+3w}}.
\label{GTexp}
\end{eqnarray}
We have checked that Eqs. (\ref{eomUD}) and (\ref{eomP2}) are
related by (\ref{GTexp}) (ignoring higher gradients). This provides
a good consistency check on our calculation. In a similar way, one
can confirm that the solutions of Eqs. (\ref{eomUD}) and
(\ref{eomP2}) are related by (\ref{GTexp}), and we get two relations
between the integration constants in the two gauges as
\begin{eqnarray}
{Y_{ij}}_{P}&=&{Y_{ij}}_{UD},\nonumber\\
{Z_{ij}}_{P}&=&{Z_{ij}}_{UD}+\frac{(1+3w)^2}{5+3w}\left[\partial_iW\partial_jV+\partial_jW\partial_iV\right]^{TT}.
\label{ICrelations}
\end{eqnarray}
We note that the gauge-dependence of the second-order tensor perturbation, (\ref{GTexp}), affects only the amplitude of the time-dependent parts of the perturbation in Poisson or uniform-density gauges, and leaves the growing mode, $Y_{ij}$, unaffected on large scales.

\section{Matching conditions}
\label{Sect:matching}

In this section, we shall derive the matching conditions for the metric and the extrinsic curvature up to second order. These matching conditions are applicable on large scales and when there is an abrupt change in the expansion rate at the matching surface. We will first consider the case where this transition happens at a specific energy density, as it does for adiabatic perturbations. We will then discuss the more general situation, where the transition is not determined by the density but by some other field $\chi$. We call this case the non-adiabatic matching.
(See section~9 of Ref.\cite{Malik:2008im} for the definition of non-adiabatic perturbations of multiple fluids or fields.)
In the final subsection, we will apply these matching conditions to the second-order tensor perturbations coming from first-order scalar-perturbations.

Matching the induced metric and extrinsic curvature of a uniform-density or uniform-$\chi$ hypersurface specifies a physical model for the matching, but does not specify the gauge. We will choose to work in a gauge in which the matching surface coincides with a constant-$\eta$ hypersurface. This is not a physical restriction; it is simply a choice of gauge, reflecting our freedom to choose the time-slicing.

The metric of the constant $\eta$ hypersurfaces (in the perturbed spacetime) is
\begin{equation}
q_{ij}=g_{ij}=a(\eta)^2\left[\left(1-2\sum_{r=1}^{+\infty}
\frac{1}{r!}\C^{(r)}\right)\delta_{ij}+\sum_{r=1}^{+\infty}
\frac{1}{r!}C^{(r)}_{ij}\right].\label{induced}
\end{equation}
The unit time-like vector field orthogonal to these surfaces is
$N^\mu=(N^0,N^i)$ with the normalization $N^\mu N_\mu=-1$. $q_{ij}$
is given by $q_{\mu\nu}=g_{\mu\nu}+N_\mu N_\nu$ which together with
equation (\ref{induced}) implies $N_\mu=(N_0,\mathbf{0})$.
The inverse induced metric is $q^{\mu\nu}=g^{\mu\nu}+N^\mu N^\nu$
and it gives
\begin{equation}
q^{ij(0)}=g^{ij(0)}, \quad q^{ij(1)}=g^{ij(1)}, \quad
q^{ij(2)}=g^{ij(2)}+a^{-2}B^{i(1)}B^{j(1)}.
\end{equation}
The extrinsic curvature of the constant $\eta$ hypersurfaces is
\begin{equation}
K_{\mu\nu}=\frac{1}{2}\pounds_{N^\lambda}q_{\mu\nu}
=\frac{1}{2}\left(q_{\mu\nu,\sigma}N^\sigma+N^\sigma_{,\mu}q_{\sigma\nu}
+N^\sigma_{,\nu}q_{\mu\sigma}\right).
\end{equation}
Defining the trace and the traceless parts of the extrinsic
curvature like
\begin{equation}
K=q^{ij}K_{ij}, \quad \tilde K_{ij}=K_{ij}-\frac{1}{3}q_{ij}K,
\end{equation}
the junction conditions are
\begin{equation}
 \label{JCs}
\left[q_{ij}\right]_-^+=0, \quad \left[K_{ij}\right]_-^+=0,
\end{equation}
which are equivalent to
\begin{equation}
\left[q_{ij}\right]_-^+=0, \quad \left[\tilde K_{ij}\right]_-^+=0,
\quad \left[K\right]_-^+=0.
\end{equation}

\subsection{Background}

In the background,
\begin{equation}
N^{0(0)}=\mp a^{-1}, \quad N^{i(0)}=0, \quad N_0^{(0)}=\pm a,
\end{equation}
and \begin{equation} K_{ij}^{(0)}=\mp a'\delta_{ij}.
\end{equation}
Therefore
\begin{equation}
K^{(0)}=\mp3\frac{\mathcal{H}}{a}, \quad \tilde K_{ij}^{(0)}=0.
\end{equation}
Continuity of the induced metric and its extrinsic curvature,
Eq.~(\ref{JCs}), at zeroth-order thus requires that $a$ and $a'$ are
continuous, i.e.,
 \begin{equation}
 \left[ a \right]_-^+=0, \quad
 \left[ \H \right]_-^+=0 \,.
 \label{JCbackground}
 \end{equation}
These matching conditions can be used to give the initial conditions
for the scale factor and its derivative after a sudden transition in
the equation of state from a $w_-$ era to a $w_+$ era (such as from
inflation to a radiation domination era). The solution of the
background Einstein equations for the conformal Hubble rate is
 \begin{equation}
 \mathcal{H} = \left(\frac{1+3w}{2}\eta+C_1\right)^{-1},
 \end{equation}
where $C_1$ is an integration constant.
From the continuity of the scale factor and its first derivative we
can obtain
\begin{eqnarray}
\mathcal{H}&=&\left(\frac{1+3w_-}{2}\eta+{C_1}_-\right)^{-1}\,\,\,\,\,\mathrm{for}\,\,\,\,\,\eta<\eta_*,\nonumber\\
\mathcal{H}&=&\left(\frac{1+3w_+}{2}\eta+{C_1}_+\right)^{-1}\,\,\,\,\,\mathrm{for}\,\,\,\,\,\eta_*<\eta,
\end{eqnarray}
where the constant ${C_1}_+$ is
\begin{equation}
{C_1}_+={C_1}_-+\frac{3}{2}\eta_*\left(w_--w_+\right).
\end{equation}
We can write the previous equations as
\begin{eqnarray}
\mathcal{H}&=&\left(\frac{1+3w_-}{2}\eta_-\right)^{-1}\,\,\,\,\,\mathrm{for}\,\,\,\,\,\eta_-<{\eta_-}_*,\nonumber\\
\mathcal{H}&=&\left(\frac{1+3w_+}{2}\eta_+\right)^{-1}\,\,\,\,\,\mathrm{for}\,\,\,\,\,{\eta_+}_*<\eta_+,
\end{eqnarray}
where the new time variables $\eta_\pm$ are
\begin{equation}
\eta_\pm=\eta+\frac{2}{1+3w_\pm}{C_1}_\pm,
\end{equation}
and the transition times in the new variables are
\begin{equation}
{\eta_-}_*=\eta_*+\frac{2}{1+3w_-}{C_1}_-,\quad
{\eta_+}_*=\frac{1+3w_-}{1+3w_+}\eta_*+\frac{2}{1+3w_+}{C_1}_-.
\end{equation}
The subscripts $+$ ($-$) denote that the quantity should be
evaluated at a time after (before) the transition time $\eta_*$.

\subsection{At first order}

The first-order matching conditions have been calculated previously
in Refs.~\cite{Deruelle:1995kd, Martin:1997zd, Copeland:2006tn}.

Continuity of the induced metric at first order leads to the
conservation of the spatial metric perturbations $C^{(1)}$ and
$C^{(1)}_{ij}$, or equivalently
\begin{eqnarray}
 \label{JCC1}
 \left[ C^{(1)} \right]_-^+ = 0 \,,\qquad
 \left[ E^{(1)} \right]_-^+ = 0 \,, \nonumber\\
 \left[ F_i^{(1)} \right]_-^+ = 0 \,,\qquad
 \left[ h_{ij}^{(1)} \right]_-^+ = 0 \,.
 \end{eqnarray}
The first-order correction to the ortho-normal vector field is
\begin{equation}
N^{0(1)}=\pm a^{-1}\A^{(1)}, \quad N^{i(1)}=\pm a^{-1}B^{i(1)},
\quad N_0^{(1)}=\pm a\A^{(1)},
\end{equation}
and the extrinsic curvature is then
\begin{equation}
K_{ij}^{(1)}=\pm\frac{a}{2}\left[2\delta_{ij}\left(2\mathcal{H}\C^{(1)}+\mathcal{H}\A^{(1)}+{\C^{(1)}}'\right)
+B_{j,i}^{(1)}+B_{i,j}^{(1)}-2\mathcal{H}C_{ij}^{(1)}-{C^{(1)}_{ij}}'
\right],
\end{equation}
or equivalently
\begin{equation}
K^{(1)}=\pm
a^{-1}\left[3\left(\mathcal{H}\A^{(1)}+{\C^{(1)}}'\right)+\partial^2B^{(1)}\right],
\quad \tilde K_{ij}^{(1)}=\pm
a\left(B_{(i,j)}^{(1)}-\frac{1}{3}\delta_{ij}\partial^2B^{(1)}-\frac{1}{2}{C_{ij}^{(1)}}'\right).
\end{equation}
The junction condition (\ref{JCs}) for the extrinsic curvature gives
 \begin{eqnarray}
\left[3\left(\mathcal{H}\A^{(1)}+{\C^{(1)}}'\right)+\partial^2B^{(1)}\right]_-^+=0,\label{JC1}\\
\left[B^{(1)}-{E^{(1)}}'\right]_-^+=0,\label{omegaparallel}\\
\left[S_j^{(1)}-{F_j^{(1)}}'\right]_-^+=0,\\
\left[{h^{(1)}_{ij}}'\right]_-^+=0.
\end{eqnarray}
Note that Eq.~(\ref{JC1}) combined with Eqs.~(\ref{JCC1}) and the
Einstein equations, Eq.~(\ref{scalarEinstein}), enforces continuity of energy density across the
hypersurface
\begin{equation}
\label{rhoJC}
 \left[ \delta\rho^{(1)} \right]_-^+ = 0\,.
 \end{equation}

\subsubsection{Adiabatic matching}

For adiabatic perturbations on large scales a sudden transition in
the equation of state must occur at a specific density. Thus we
apply the matching conditions across uniform-density hypersurface
\cite{Deruelle:1995kd}.
We can use the freedom in the choice of spatial coordinates on the
matching hypersurface to set $E^{(1)}=0$. In this case the matching
conditions (\ref{JCC1}) and (\ref{omegaparallel}) across a
uniform-density hypersurface reduce to
\begin{equation}
 \label{adiabaticJCs}
 \left[ \zeta \right]_-^+ = 0 \,, \qquad
 \left[ \Psi \right]_-^+ =0 \,.
 \end{equation}
Note that the remaining junction condition for scalar perturbations,
Eq.~(\ref{JC1}), reduces to $[\zeta^{\prime}]_-^+=0$ on large scales
which is automatically satisfied for adiabatic perturbations since
$\zeta^{\prime}=0$.
Taking the growing mode for adiabatic perturbations on large scales
before the transition
 \begin{equation}
\Psi_- = -\frac{3+3w_-}{5+3w_-}{V}_- = {\rm const}\label{Psi-}
\,,
 \end{equation}
then the junction conditions (\ref{adiabaticJCs}), together with the
general solution for constant $w$, Eq. (\ref{Phisol}), can be used to
determine the solution of $\Psi$ after a sudden transition as
 \begin{eqnarray}
\Psi_+&=&-\frac{3+3w_+}{5+3w_+}{V}_+ +
W_+\eta_+^{-\frac{5+3w_+}{1+3w_+}},\label{Psi+}
 \end{eqnarray}
where $V_+=V_-=\zeta_-$ and the amplitude of the decaying mode, $W_+$, is given by
\begin{equation}
 \label{Wplus}
 W_+ =
  \frac{3+3w_+}{5+3w_+}{\eta_+}_*^{\frac{5+3w_+}{1+3w_+}}\left(1-\frac{3+3w_-}{5+3w_-}\frac{5+3w_+}{3+3w_+}\right){V}_-.
\end{equation}

\subsubsection{Non-adiabatic matching}

To consider a more general matching condition we will consider the
case where the transition surface is given by a fixed value of some
field, $\chi$.
The scalar metric perturbations on a uniform-$\chi$ hypersurface can
be given in terms of the metric perturbations and $\chi$-field
perturbations in an arbitrary gauge as
\begin{eqnarray}
 -\zeta_\chi = C_\chi^{(1)} + \frac13 \partial^2 E_\chi^{(1)}
  &=& C^{(1)} + \frac13 \partial^2 E^{(1)} + \H \frac{\delta\chi^{(1)}}{\chi^{(0)\prime}}
 \,,\\
 B_\chi^{(1)} - E_\chi^{(1)\prime} &=& B^{(1)} - E^{(1)\prime} +
 \frac{\delta\chi^{(1)}}{\chi^{(0)\prime}} \,.
 \end{eqnarray}
Departures from adiabaticity are characterised by perturbations of
the $\chi$-field with respect to the total density
\begin{equation}
 \S_\chi \equiv \H \left( \frac{\delta\rho^{(1)}}{\rho^{(0)\prime}}
  - \frac{\delta\chi^{(1)}}{\chi^{(0)\prime}} \right)
  \,.
 \end{equation}
We will recover the adiabatic matching conditions when $\S_\chi=0$
and $\chi$ is unperturbed on uniform-density hypersurfaces, so that
the uniform-$\chi$ hypersurface coincides with a uniform-density
hypersurface.

The usual gauge-invariant curvature perturbations can then be given
as
\begin{eqnarray}
 \zeta &=& \zeta_\chi - \S_\chi \,,\\
 \Psi &=& -\zeta_\chi -\H \left( B_\chi^{(1)} - E_\chi^{(1)\prime}
 \right) \,.
 \end{eqnarray}

The junction conditions for scalar metric perturbations (\ref{JCC1})
and (\ref{omegaparallel}) require
\begin{equation}
 \label{chiJCs}
 \left[ \zeta_\chi \right]_-^+ = 0 \,, \quad
 \left[ B_\chi^{(1)} - E_\chi^{(1)\prime} \right]_-^+ =0 \,,
 \end{equation}
while Eq.~(\ref{JC1}) enforces energy conservation across the
hypersurface, Eq.~(\ref{rhoJC}) and thus
\begin{equation}
 \left[ \S_\chi \right]_-^+ = \left( \frac{1+w_-}{1+w_+} - 1 \right)
 \S_{\chi-} \,.
 \end{equation}

Expressing Eqs.~(\ref{chiJCs}) in terms of the usual gauge-invariant
curvature perturbations we then have
\begin{equation}
 \left[ \zeta \right]_-^+ = \left( 1 - \frac{1+w_-}{1+w_+}\right)
 \S_{\chi-} \,,
 \qquad
 \label{chiJCPsi}
  \left[ \Psi^{(1)} \right]_-^+ =0 \,.
 \end{equation}

Analogously to the previous subsection, these junction conditions can be used to evolve the solution of $\Psi$ across the transition. The solution is given by Eqs. (\ref{Psi-}) and (\ref{Psi+}), where now the integration constants after the transition are given by
\begin{equation}
 \label{chiJCW}
V_+ =
\zeta_{\chi-}+\frac{1+w_-}{1+w_+}\left(V_--\zeta_{\chi-}\right) ,
\quad W_+ =
\left(\frac{3+3w_+}{5+3w_+}V_+-\frac{3+3w_-}{5+3w_-}V_-\right)\eta_*^\frac{5+3w_+}{1+3w_+}.
\end{equation}

\subsection{At second order}

Continuity of the induced metric at second order leads to the
obvious extension of the first-order junction conditions
(\ref{JCC1}) for the spatial metric perturbations $C^{(2)}$ and
$C^{(2)}_{ij}$, or equivalently
\begin{eqnarray}
 \label{JCC2}
 \left[ C^{(2)} \right]_-^+ = 0 \,,\qquad
 \left[ E^{(2)} \right]_-^+ = 0 \,, \nonumber\\
 \left[ F_i^{(2)} \right]_-^+ = 0 \,,\qquad
 \left[ h_{ij}^{(2)} \right]_-^+ = 0 \,.
 \end{eqnarray}
At second order, in the ortho-normal vector field we have
 \begin{eqnarray}
N^{0(2)}&=&\pm
\frac{a^{-1}}{2}\left[B_i^{(1)}B^{i(1)}-3\left(\A^{(1)}\right)^2+\A^{(2)}\right],
\quad N_0^{(2)}=\pm
\frac{a}{2}\left[B_i^{(1)}B^{i(1)}-\left(\A^{(1)}\right)^2+\A^{(2)}\right],
 \\
N^{i(2)}&=&\pm
a^{-1}\left(-\A^{(1)}B^{i(1)}+2\C^{(1)}B^{i(1)}-B_j^{(1)}C^{ij(1)}+\frac{B^{i(2)}}{2}\right).
 \end{eqnarray}
The extrinsic curvature tensor is
\begin{eqnarray}
K_{ij}^{(2)}&=&\pm
a\delta_{ij}\left[\mathcal{H}\C^{(2)}+\frac{1}{2}{\C^{(2)}}'+\frac{1}{2}\mathcal{H}\A^{(2)}-2\mathcal{H}\A^{(1)}\C^{(1)}
-\A^{(1)}{\C^{(1)}}'+\frac{1}{2}\mathcal{H}\left(B_k^{(1)}B^{k(1)}-3\left(\A^{(1)}\right)^2\right)
-B^{k(1)}\C^{(1)}_{,k}\right]\nonumber
\\&\pm&a\Bigg[-\frac{\mathcal{H}}{2}C_{ij}^{(2)}-\frac{1}{4}{C_{ij}^{(2)}}'+\mathcal{H}\A^{(1)}C_{ij}^{(1)}
+\frac{1}{2}\A^{(1)}{C_{ij}^{(1)}}'+\frac{1}{2}B^{k(1)}C_{ij,k}^{(1)}-\A^{(1)}B_{(j,i)}^{(1)}+2\C^{(1)}_{,(i}B_{j)}^{(1)}
-B_k^{(1)}C_{(j,i)}^{k(1)}+\frac{B_{(j,i)}^{(2)}}{2}
\Bigg].\nonumber\\
\end{eqnarray}
The trace and the traceless part read
\begin{eqnarray}
 \label{K2}
K^{(2)}&=&\pm\frac{1}{2a}\Big[3\left(\mathcal{H}\A^{(2)}+{\C^{(2)}}'\right)+B_i^{,i(2)}+2\left(3{\C^{(1)}}'+B^{,j(1)}_j\right)\left(2\C^{(1)}-\A^{(1)}\right)
-9\mathcal{H}\left(\A^{(1)}\right)^2\nonumber
\\&&-2\C^{,i(1)}B_i^{(1)}+3\mathcal{H}B_i^{(1)}B^{i(1)}
-2C^{ij(1)}B_{j,i}^{(1)}-2B^{i(1)}C^{,j(1)}_{ij}+C^{ij(1)}{C_{ij}^{(1)}}'
\Big],\nonumber
\\\tilde K_{ij}^{(2)}&=&\pm a \Bigg(-\frac{{C_{ij}^{(2)}}'}{4}+\frac{1}{2}B_{(i,j)}^{(2)}+\frac{1}{2}\A^{(1)}{C_{ij}^{(1)}}'+\frac{1}{2}B^{k(1)}C_{ij,k}^{(1)}
-{\C^{(1)}}'C_{ij}^{(1)}-\frac{1}{3}\partial^2B^{(1)}C_{ij}^{(1)}\nonumber
\\&&-\A^{(1)}B_{(i,j)}^{(1)}+2\C^{(1)}_{,(i}B_{j)}^{(1)}
-B^{(1)}_{k,(i}C^{k(1)}_{j)}-B_k^{(1)}C^{k(1)}_{(i,j)}+B^{k(1)}_{,(i}C^{(1)}_{j)k}
\Bigg)\nonumber
\\&&\mp a\frac{\delta_{ij}}{3}\Bigg(\frac{B_k^{k(2)}}{2}-\A^{(1)}B_k^{,k(1)}+2\C^{,k(1)}B_k^{(1)}-B^{k,l(1)}C_{kl}^{(1)}
-B_k^{(1)}C^{k,l(1)}_l+\frac{1}{2}C^{kl(1)}{C_{kl}^{(1)}}'\Bigg).
\end{eqnarray}
The remaining matching conditions are then
\begin{equation}
 \label{JCK2}
\left[\tilde K_{ij}^{(2)}\right]_-^+=0, \quad
\left[K^{(2)}\right]_-^+=0.
\end{equation}

\subsection{Matching conditions for second-order tensors from first-order scalars}

\subsubsection{Adiabatic matching}

For adiabatic perturbations we match on a uniform-density
hypersurface, $\delta\rho=0$.
Matching the induced metric on the transition surface then implies,
from Eq.~(\ref{JCC2}), that
\begin{equation}
 \label{hUDJCad}
 \left[ {h_{ij}^{(2)}}_{UD} \right]_-^+ = 0 \,.
\end{equation}
Setting $E^{(1)}=0$ and neglecting first-order vector and tensor
perturbations, so that $C^{(1)}_{ij}=0$, the matching condition
(\ref{JCK2}) for the transverse and traceless part of the extrinsic
curvature (\ref{K2}) gives
\begin{equation}
\Bigg[{h_{ij}^{(2)}}'_{UD}
\Bigg]_-^+=-4\Bigg[\left(\A^{(1)}_{UD}B_{UD,ij}^{(1)}
 +2\zeta_{,(i}B_{UD,j)}^{(1)} \right)^{TT}
\Bigg]_-^+.\label{JCTTP2}
\end{equation}

Continuity of $\zeta$ and $B^{(1)}_{UD}$ across the matching
hypersurface is enforced by the first-order scalar matching
conditions (\ref{JCC1}) and~(\ref{omegaparallel}). In addition, in
the uniform-density gauge, from (\ref{JC1}) we have
\begin{equation}
\left[ \A^{(1)}_{UD} \right]_-^+
 =
 \left[\frac{\zeta'}{\mathcal{H}}\right]_-^+
 \,.\label{JC2}
\end{equation}
Thus for adiabatic perturbations on large scales, for which
$\zeta'=0$, we find
\begin{equation}
 \label{hprimeUDJCad}
\Bigg[{h_{ij}^{(2)}}'_{UD} \Bigg]_-^+ = 0 \,.
\end{equation}
Thus the tensor metric perturbation and its first derivative are
continuous in the uniform-density gauge for adiabatic perturbations
on large scales.

The second-order gauge transformation from the uniform-density gauge to the Poisson gauge yields
\begin{eqnarray}
 \label{hPtoUD}
{h_{ij}^{(2)}}_{UD} = {h_{ij}^{(2)}}_{P} - 2 \left(
B^{(1)}_{UD,(i}B^{(1)}_{UD,j)} \right)^{TT} \,,\\
{h_{ij}^{(2)\prime}}_{UD} = {h_{ij}^{(2)\prime}}_{P} - 4 \left(
B^{(1)}_{UD,(i}B^{(1)\prime}_{UD,j)} \right)^{TT} \,.
\end{eqnarray}
For adiabatic perturbations on large scales, we have
\begin{equation}
 B^{(1)\prime}_{UD} =  \zeta + 2\Psi  \,.
\end{equation}
Thus from Eqs.~(\ref{hPtoUD}) and the continuity of
$B^{(1)}_{UD}$, $\zeta$ and $\Psi$ across the matching surface, we have
\begin{equation}
 \label{AdiabaticJChP}
 \left[ {h_{ij}^{(2)}}_{P} \right]_-^+ = 0 \,,
  \qquad
 \left[ {h_{ij}^{(2)\prime}}_{P} \right]_-^+ = 0 \,.
\end{equation}
The tensor metric perturbation ${h^{(2)}_{ij}}$ and its time
derivative are thus continuous across the matching surface in either
uniform-density or Poisson gauge for adiabatic perturbations on
large scales.

\subsubsection{Non-adiabatic matching}

In this subsection, we will consider the non-adiabatic case where the transition surface is determined by a fixed value of the field
$\chi$. Therefore we match on a uniform-$\chi$ hypersurface which,
for non-adiabatic perturbations, need not coincide with a uniform-density hypersurface.

Continuity of the induced metric on the uniform-$\chi$ hypersurface
implies
\begin{equation}
 \label{JChchi}
\left[ {h_{ij}^{(2)}}_{\chi} \right]_-^+ = 0.
\end{equation}
Because we choose $E^{(1)}_\chi=0$, Eq.~(\ref{chiJCs}) implies that
$B^{(1)}_\chi$ is continuous across the transition surface. This
fact simplifies the matching conditions for the extrinsic curvature,
Eqs.~(\ref{K2}) and~(\ref{JCK2}), to give
\begin{equation}
 \label{JChchiprime}
 \left[ {h_{ij}^{(2)\prime}}_{\chi} \right]_-^+ = -4\left[\left(A^{(1)}_\chi B_{\chi,ij}^{(1)}\right)^{TT}\right]_-^+  \,.
\end{equation}
The second-order gauge transformation for the tensor metric
perturbation from a general gauge to the uniform-$\chi$ gauge is
given by Eq.~(\ref{GT}), where from the first-order gauge
transformation for a scalar (\ref{tilderho}) gives
\begin{equation}
 \alpha^{(1)}_\chi = -\frac{\delta\chi^{(1)}}{\chi^{(0)}} .
\end{equation}
Thus the matching condition (\ref{JChchi}) can be written in terms
of the metric perturbations in an arbitrary gauge as
\begin{equation}
\left[h^{(2)}_{ij}\right]_-^+  = 2\left[\left(2B^{(1)}-\alpha^{(1)}_{\chi} \right)
\alpha^{(1)}_{\chi,ij}\right]^{+TT}_-.\label{JCInducedMetricNon-A}
\end{equation}
Matching the derivative (\ref{JChchiprime}) in an arbitrary gauge we
obtain
\begin{equation}
\left[h^{(2)'}_{ij}\right]_-^+  = 4\left[B^{(1)'}\alpha^{(1)}_{\chi,ij}-\left(B^{(1)}-\alpha^{(1)}_\chi\right)\left(A_{,ij}^{(1)}+\mathcal{H}\alpha^{(1)}_{\chi,ij}\right)\right]^{+TT}_-.
\label{JCExtrinsicCurvatureNon-A}
\end{equation}
In the Poisson gauge $B_P^{(1)}=0$, $A_P^{(1)}=\Psi$ and we have
\begin{eqnarray}
\left[h^{(2)}_{ij~P}\right]_-^+  &=& - 2\left[ \alpha^{(1)P}_{\chi}
\alpha^{(1)P}_{\chi,ij}\right]^{+TT}_-,\label{JCInducedMetricNon-AP}
 \\
\left[h^{(2)'}_{ij~P}\right]_-^+  &=& -
4\left[\alpha^{(1)P}_\chi\left(\Psi_{,ij}+\mathcal{H}\alpha^{(1)P}_{\chi,ij}\right)\right]^{+TT}_-,
\label{JCExtrinsicCurvatureNon-AP}
\end{eqnarray}
where
\begin{equation}
\alpha^{(1)P}_\chi = - \frac{\delta\chi_P^{(1)}}{\chi^{(0)\prime}}
 = \frac{\zeta_\chi+\Psi}{\mathcal{H}}.
\end{equation}
Equations (\ref{chiJCs}), (\ref{chiJCPsi}) and (\ref{JCbackground})
show that $\zeta_\chi$, $\Psi$ and $\mathcal{H}$ are continuous, and
thus $\alpha^{(1)P}_\chi$ is also continuous. Thus Eqs.
(\ref{JCInducedMetricNon-AP}) and~(\ref{JCExtrinsicCurvatureNon-AP})
show that the tensor perturbation,
${h^{(2)}_{ij}}_P$, and its derivative, ${h^{(2)'}_{ij}}_P$, are
continuous in the Poisson gauge, even in the non-adiabatic matching
case.

In the uniform-density gauge the situation is rather different.
$B_{UD}^{(1)}$ is given by Eq.~(\ref{scalarshiftUD}) and
$A_{UD}^{(1)}=\zeta'/\mathcal{H}$ on large scales, and we have
\begin{equation}
\alpha^{(1)UD}_\chi = -
\frac{\delta\chi_{UD}^{(1)}}{\chi^{(0)\prime}}
 = \frac{\zeta_\chi-\zeta}{\mathcal{H}}.
\end{equation}
Allowing for the fact that $\zeta_\chi$ and $\Psi$ are continuous,
the matching conditions~(\ref{JCInducedMetricNon-A})
and~(\ref{JCExtrinsicCurvatureNon-A}) reduce to
\begin{eqnarray}
\left[h^{(2)}_{ij~UD}\right]_-^+
 &=&
 \frac{2}{\mathcal{H}^2} \left[ (\zeta+2\Psi) \zeta_{,ij} \right]^{+TT}_-,
\label{JCInducedMetricNon-AUD}
 \\
\left[h^{(2)'}_{ij~UD}\right]_-^+
 &=&
 \frac{4}{\mathcal{H}^2} \left[ (\zeta'-\mathcal{H}\zeta-2\mathcal{H}\Psi) (\zeta+\Psi)_{,ij} \right]^{+TT}_-.
\label{JCExtrinsicCurvatureNon-AUD}
\end{eqnarray}
If $\zeta$ is not continuous across the transition then the tensor
perturbation in the uniform-density gauge, and its time derivative,
will not be continuous.

Assuming the evolution is piece-wise adiabatic (that is, $\zeta'=0$
before and after the transition) and using Eq.~(\ref{chiJCPsi}) for
the jump in $\zeta$ due to the non-adiabatic transition, we have
\begin{eqnarray}
 \label{JChUDnad}
\left[{h^{(2)}_{ij}}_{UD}\right]_-^+ &=&
 \frac{2}{\mathcal{H}^2}
 \left( \frac{w_+-w_-}{1+w_+} \right) \left(
 \left\{ \frac{w_+-w_-}{1+w_+} \mathcal{S}_{\chi-}+\frac{4}{5+3w_-}\zeta_-\right\}
 \mathcal{S}_{\chi-,ij}
 \right)^{TT} \,,\\
  \label{JChprimeUDnad}
\left[{h^{(2)\prime}_{ij}}_{UD}\right]_-^+ &=&
 - \frac{4}{\mathcal{H}}
 \left( \frac{w_+-w_-}{1+w_+} \right) \left(
 \left\{ \frac{w_+-w_-}{1+w_+} \mathcal{S}_{\chi-}+\frac{1-3w_-}{5+3w_-}\zeta_-\right\}
 \mathcal{S}_{\chi-,ij}
 \right)^{TT} \,.
\end{eqnarray}


\section{Gravitational waves from inflationary perturbations on large scales}

As an application of the matching conditions for tensor
perturbations we will consider the generation of gravitational waves
on super-Hubble scales after the end of inflation. During slow-roll
inflation we have $\epsilon\ll 1$ where the slow-roll parameter is
defined as $\epsilon\equiv-{\dot H}/{H^2}=3(1+w)/2$.

Using the inflationary solution to set the initial conditions for
the tensor metric perturbation on super-Hubble scales we can use the
solution (\ref{solP}) with the equation of state
$w_-=-1+(2/3)\epsilon$. During inflation the conformal time $\eta$
decreases, and we neglect the decaying solutions proportional to
${Z_{ij}}_{P-}$ and $W_{-}$. Then the large scale solution in the
Poisson gauge, Eq.~(\ref{solP}), is
\begin{equation}
{h_{ij}^{(2)}}_{P-} =
{Y_{ij}}_{P-}-4\epsilon\eta_-^2\left[\partial_iV_-\partial_jV_-\right]^{TT}\,.
\end{equation}
The constant solution on large scales, ${Y_{ij}}_{P-}$, is the usual
free part of the gravitational wave solution whose amplitude is
determined by the quantum vacuum on small scales, assuming inflation
is sufficiently long-lived.
We are specifically interested in the production of second-order
tensor perturbations from first-order scalar perturbations, $\zeta$
or $\Psi$, during inflation. Hence we choose to set ${Y_{ij}}_{P-}$
equal to zero and study the solution
\begin{equation}
 \label{hPinflation}
{h_{ij}^{(2)}}_{P-} =
-\frac{4\epsilon}{\mathcal{H}^2}\left[\partial_iV_-\partial_jV_-\right]^{TT},
\quad
 {h_{ij}^{(2)'}}_{P-} =
\frac{8\epsilon}{\mathcal{H}}\left[\partial_iV_-\partial_jV_-\right]^{TT}.
\end{equation}
The tensor metric perturbations produced by the first-order scalar
metric perturbations during inflation are thus suppressed in the
Poisson gauge, and vanish in the limit $\epsilon\to0$.

One can use the gauge transformation equation (\ref{GTexp}) during
inflation to find the inflationary initial conditions in the
uniform-density gauge from the Poisson gauge initial conditions as
\begin{equation}
 \label{hUDinflation}
{h_{ij}^{(2)}}_{UD-}=-\frac{2}{\mathcal{H}^2}\left[\partial_iV_-\partial_jV_-\right]^{TT},
\quad
{h_{ij}^{(2)}}_{UD-}'=\frac{4}{\mathcal{H}}\left(1-\epsilon\right)\left[\partial_iV_-\partial_jV_-\right]^{TT}.
\end{equation}
We see that the tensor metric perturbations during inflation
produced from first-order metric perturbations are not slow-roll
suppressed in the uniform-density gauge, in contrast to the Poisson
gauge result~(\ref{hPinflation}).

Equations~(\ref{hPinflation}) or~(\ref{hUDinflation}) together with
the matching conditions obtained in Section~\ref{Sect:matching} then
set the initial conditions on large scales at the start of the
radiation era that follows the inflationary period.

\subsection{Adiabatic matching}

First-order scalar metric perturbations can be calculated in any
gauge, but the curvature perturbation in the uniform-density gauge,
$\zeta$, remains constant after inflation for adiabatic
perturbations and on large scales \cite{Wands:2000dp}, and hence we
have $V_+=V_-$. The curvature perturbation in the longitudinal
gauge, $\Psi$, is also continuous for adiabatic perturbations, as
shown by the matching conditions Eq.~(\ref{adiabaticJCs}), but the
continuity of $\zeta$ implies from Eq.~(\ref{zetaeq}) that $\Psi'$
is discontinuous at a sudden change in the equation of state. Hence
$\Psi$ becomes time-dependent after inflation.

The adiabatic matching conditions for the scalar metric
perturbations (\ref{adiabaticJCs}) determine the amplitude of the
decaying mode (\ref{Wplus}) in a radiation era, with equation of
state $w_+=1/3$, following slow-roll inflation
\begin{eqnarray}
 \label{Wplusad}
W_+&=&\frac{2}{3}\left(1-\frac{3}{2}\epsilon\right)\eta_{+*}^3V_-.
\end{eqnarray}
During the radiation era the general solution for tensor metric
perturbations in the Poisson gauge, Eq.~(\ref{solP}), reduces to
\begin{equation}
 \label{hPrad}
{h_{ij}^{(2)}}_{P+}={Y_{ij}}_{P+}+{Z_{ij}}_{P+}\eta_+^{-1}
+\frac{8}{9}\left[\partial_iV_+\partial_jV_+\right]^{TT}\eta_+^2+2\left[\partial_iW_+\partial_jW_+\right]^{TT}\eta_+^{-4}.
\end{equation}
Using the adiabatic matching conditions for the second-order tensor
perturbation in the Poisson gauge~(\ref{AdiabaticJChP}), which
require that both the tensor perturbation and its first derivative
are continuous at the end of inflation, we find that during the
radiation era following inflation the integration constants are
given by
\begin{equation}
 \label{YZPinf}
{Y_{ij}}_{P+}=-4\epsilon\eta_{+*}^2\left[\partial_iV_-\partial_jV_-\right]^{TT},
\quad
{Z_{ij}}_{P+}=-\frac{16}{9}\left(1-\frac{3}{2}\epsilon\right)\eta_{+*}^3\left[\partial_iV_-\partial_jV_-\right]^{TT}.
\end{equation}

The general solution for the tensor perturbation in the
uniform-density gauge (\ref{hUD}) for $w_+=1/3$ becomes
\begin{equation}
 \label{hUDrad}
{h_{ij}^{(2)}}_{UD+} =
{Y_{ij}}_{UD+} + {Z_{ij}}_{UD+}\eta_+^{-1} + \frac23
\left[\partial_iV_+\partial_jV_+\right]^{TT}\eta_+^2 \,.
\end{equation}
Either from the adiabatic matching conditions for the tensor
perturbations in the uniform-density gauge (\ref{hUDJCad}) and
(\ref{hprimeUDJCad}), or from the gauge transformation
Eqs.~(\ref{ICrelations}) of the Poisson gauge results
(\ref{YZPinf}), one can obtain the uniform-density integration
constants as
\begin{eqnarray}
{Y_{ij}}_{UD+}=-4\epsilon\eta_{+*}^2\left[\partial_iV_-\partial_jV_-\right]^{TT}, \quad
{Z_{ij}}_{UD+}=-\frac{8}{3}\left(1-\frac{3}{2}\epsilon\right)\eta_{+*}^3\left[\partial_iV_-\partial_jV_-\right]^{TT}.
\end{eqnarray}
Thus the full solution for the induced tensor perturbations,
$h_{ij}^{(2)}$, on large scales during a radiation era after
inflation, is given in the Poisson gauge (\ref{hPrad}) and
uniform-density gauge (\ref{hUDrad}) by
\begin{equation}
 \label{hPinfplus}
{h_{ij}^{(2)}}_{P+} = \left[- 4\epsilon\eta^2_{+*} +
\frac{8}{9}\eta_+^2 -
\frac{16}{9}\left(1-\frac{3}{2}\epsilon\right)\eta_{+*}^3\eta_+^{-1}
+ \frac{8}{9}\left(1-3\epsilon\right)\eta_{+*}^6{\eta_+}^{-4}\right]
\left[\partial_iV_-\partial_jV_-\right]^{TT},
\end{equation}
\begin{equation}
 \label{hUDinfplus}
{h_{ij}^{(2)}}_{UD+} = \left[ - 4\epsilon\eta^2_{+*} +
\frac{2}{3}\eta_+^2 -
\frac{8}{3}\left(1-\frac{3}{2}\epsilon\right)\eta_{+*}^3\eta_+^{-1}\right]
\left[\partial_iV_-\partial_jV_-\right]^{TT}.
\end{equation}
We see that although the time-dependent parts of the tensor
perturbation in the two gauges are different, the constant mode
after inflation in both gauges is the same and it is slow-roll
suppressed, ${Y_{ij}}_{P+}={Y_{ij}}_{UD+}={\cal O}(\epsilon)$, even
in the uniform-density gauge where the tensor perturbation is not
suppressed during slow-roll inflation.

\subsection{Non-adiabatic matching}

In the presence of non-adiabatic perturbations the scalar curvature
perturbation $\zeta$ needs no longer be constant on large scales and may be
discontinuous, as shown in Eq.~(\ref{chiJCPsi}).
As a simple model of non-adiabatic perturbations we consider the
case where the transition from slow-roll inflation to radiation
occurs on a uniform-$\chi$ hypersurface with curvature
$\zeta_\chi\neq\zeta_-$. {}From the non-adiabatic matching condition
for scalar metric perturbations, Eq.~(\ref{chiJCW}), we obtain the
amplitude of the constant mode and the decaying mode of $\Psi$ in the
radiation era following inflation
\begin{eqnarray}
 \label{nadV}
V_+ &=& \left( 1-\frac{\epsilon}{2} \right) \zeta_{\chi-} +
\frac{\epsilon}{2} V_- \,,\\
 \label{nadW}
W_+&=&\frac{1}{3}\Big((2-\epsilon)\zeta_{\chi-}-2\epsilon
V_-\Big)\eta_{+*}^3 \,.
\end{eqnarray}
For $\zeta_\chi=V_-$ we recover the adiabatic case (\ref{Wplusad}).

The tensor perturbation in the Poisson gauge, and its time
derivative remain continuous for a non-adiabatic transition.
Matching the inflationary solution (\ref{hPinflation}) to the
radiation era solution (\ref{hPrad}) with the integration constant
$V_+$ and $Z_+$ given by Eqs. (\ref{nadV}) and (\ref{nadW})
determines the two remaining integration constants in the Poisson
gauge
\begin{eqnarray}
{Y_{ij}}_{P+}&=&4\epsilon\eta_{+*}^2\Big[\partial_iV_-\partial_jV_--2\partial_i\zeta_{\chi-}\partial_jV_-\Big]^{TT},\\
{Z_{ij}}_{P+}&=&-8\eta_{+*}^3\left[\epsilon\partial_iV_-\partial_jV_-+\frac{2}{9}(1-\epsilon)\partial_i\zeta_{\chi-}\partial_j\zeta_{\chi-}-\frac{10}{9}\epsilon\partial_i\zeta_{\chi-}\partial_jV_-\right]^{TT}.
\end{eqnarray}

The tensor metric perturbation in the uniform-density gauge can be
discontinuous, with the jump in $h_{ij}$ and its time derivative
being given by (\ref{JChUDnad}) and (\ref{JChprimeUDnad}), where the
incoming non-adiabatic perturbation is given by
\begin{equation}
S_{\chi-}=\zeta_{\chi-}-V_- \,.
\end{equation}
As a result, matching the inflationary solution (\ref{hUDinflation})
to the radiation era solution (\ref{hUDrad}) with the integration
constant $V_+$ and $Z_+$ given by Eqs. (\ref{nadV}) and (\ref{nadW})
gives the two remaining integration constants
\begin{eqnarray}
{Y_{ij}}_{UD+}&=&{Y_{ij}}_{P+},\\
{Z_{ij}}_{UD+}&=&-8\eta_{+*}^3\left[\epsilon\partial_iV_-\partial_jV_-+\frac{1}{3}(1-\epsilon)\partial_i\zeta_{\chi-}\partial_j\zeta_{\chi-}-\frac{7}{6}\epsilon\partial_i\zeta_{\chi-}\partial_jV_-\right]^{TT}.
\end{eqnarray}
Therefore the full solution for the induced tensor perturbations,
$h_{ij}^{(2)}$, on large scales during a radiation era after
inflation in this non-adiabatic model for the transition, is given
in either the Poisson gauge (\ref{hPrad}) and uniform-density gauge
(\ref{hUDrad}) by
\begin{eqnarray}
{h_{ij}^{(2)}}_{P+}&=&\frac{8}{9}\eta_+^2\left[(1-\epsilon)\partial_i\zeta_{\chi-}\partial_j\zeta_{\chi-}+\epsilon\partial_i\zeta_{\chi-}\partial_jV_-\right]^{TT}
+4\epsilon\eta_{+*}^2\left[\partial_iV_-\partial_jV_--2\partial_i\zeta_{\chi-}\partial_jV_-\right]^{TT}
\nonumber\\&&
-8\eta_{+*}^3\eta_+^{-1}\left[\epsilon\partial_iV_-\partial_jV_-+\frac{2}{9}(1-\epsilon)\partial_i\zeta_{\chi-}\partial_j\zeta_{\chi-}-\frac{10}{9}\epsilon\partial_i\zeta_{\chi-}\partial_jV_-\right]^{TT}
\nonumber\\&&
+\frac{8}{9}\eta_{+*}^6\eta_+^{-4}\left[(1-\epsilon)\partial_i\zeta_{\chi-}\partial_j\zeta_{\chi-}-2\epsilon\partial_i\zeta_{\chi-}\partial_jV_-\right]^{TT},
\end{eqnarray}
\begin{eqnarray}
{h_{ij}^{(2)}}_{UD+}&=&\frac{2}{3}\eta_+^2\left[(1-\epsilon)\partial_i\zeta_{\chi-}\partial_j\zeta_{\chi-}+\epsilon\partial_i\zeta_{\chi-}\partial_jV_-\right]^{TT}
+4\epsilon\eta_{+*}^2\left[\partial_iV_-\partial_jV_--2\partial_i\zeta_{\chi-}\partial_jV_-\right]^{TT}
\nonumber\\&&
-8\eta_{+*}^3\eta_+^{-1}\left[\epsilon\partial_iV_-\partial_jV_-+\frac{1}{3}(1-\epsilon)\partial_i\zeta_{\chi-}\partial_j\zeta_{\chi-}-\frac{7}{6}\epsilon\partial_i\zeta_{\chi-}\partial_jV_-\right]^{TT}.
\end{eqnarray}
In the adiabatic case where $\zeta_{\chi-}=V_-$ and hence
$S_{\chi-}=0$, this reduces to Eqs.~(\ref{hPinfplus})
and~(\ref{hUDinfplus}).
The presence of non-adiabatic perturbations changes the amplitude of
the decaying mode, proportional to $Z_{ij}$, and the
amplitude of the
growing mode proportional to $\eta_+^2$. But the free part of the
gravitational field, which is gauge-independent, remains constant on
super-Hubble scales and remains suppressed in the slow-roll limit
\begin{equation}
{Y_{ij}}_{UD+} = {Y_{ij}}_{P+} = {\cal O}(\epsilon)\,.
\end{equation}

\section{Conclusion}

Tensor perturbations of the FRW metric become gauge-dependent at
second and higher order \cite{Bruni:1996im}. It is however possible to
construct gauge-invariant tensor perturbations by eliminating the
gauge degrees of freedom \cite{Malik:2008im}. As examples we have
given gauge-invariant definitions of the tensor perturbation at
second order in the Poisson gauge and the uniform-density gauge, and
the (gauge-invariant) difference between the two. All of this should
be familiar from the study of gauge-dependent scalar perturbations
which are already gauge dependent at first order, and one can choose
to work in terms of the curvature perturbation in the Poisson (or
longitudinal) gauge, $\Psi$, or in the uniform-density gauge,
$\zeta$, or both.

We have focussed on the question of matching conditions for the
tensor perturbation. At first order, the Israel junction conditions
require that the tensor metric perturbation and its time-derivative
are always continuous across a spacelike hypersurface in any gauge.
But at second order we need to consider the behaviour in different
gauges. For adiabatic perturbations the matching hypersurface should
coincide with a uniform-density hypersurface. The Israel junction
conditions then imply the tensor perturbation in the uniform-density
gauge, and its time-derivative, will be continuous. For the case of
non-adiabatic matching the hypersurface need not coincide with a
uniform-density surface and so the tensor perturbation in the
uniform-density gauge may be discontinuous. However in both cases
(adiabatic or non-adiabatic) we show that the second-order tensor
perturbation in the Poisson gauge, and its time-derivative, are
continuous.

As an application we consider the generation of gravitational waves
during inflation. We give expressions for the general solution for
the tensor perturbation on large (super-Hubble) scales in both the
Poisson and uniform-density gauges when the barotropic index, $w$ is
constant. This is a good description of the evolution during the
radiation dominated era, when $w=1/3$, and during a preceding era of
slow-roll inflation, when $w=-1+(2/3)\epsilon$ and $\epsilon\ll1$ is
a slow-roll parameter. It has previously been shown \cite{Maldacena:2002vr,
Arroja:2008ga} that
the tensor metric perturbation on large-scales {\em during}
inflation is slow-roll suppressed in the Poisson gauge, but not
suppressed in the uniform-density gauge. This reflects the fact that
the scalar metric perturbations are slow-roll suppressed during
inflation in the Poisson gauge with respect to those in the
uniform-density gauge. The question is then, whether the amplitude
of gravitational waves {\em after} inflation is slow-roll
suppressed, or indeed whether large tensor perturbation can be
generated at second order by the transition to a radiation dominated
universe.

Continuity of the tensor perturbation and its first derivative is
sufficient to show that the tensor perturbation remains slow-roll
suppressed after inflation in the Poisson gauge. In the
uniform-density gauge tensor perturbations are not slow-roll
suppressed, reflecting the gauge-dependence of the second-order
tensor perturbations. However, the constant mode on large scales,
$Y_{ij}$, corresponding to the free part of the tensor perturbations
is the same in the Poisson or uniform-density gauge in the radiation
era. It is this constant mode that describes the gravitational waves
produced by inflation on large scales. In the case of non-adiabatic
perturbations this amplitude can change in the uniform-density gauge
due to non-adiabatic matching condition, but even in this case its
amplitude vanishes in the slow-roll limit $\epsilon\to 0$.

In addition to the constant mode there is a growing mode
proportional to $\eta^2$ on large scales in the radiation era. This
is not the free part of the gravitational field, i.e., it vanishes
in the absence of first-order scalar perturbations. But, like the
scalar metric perturbations during the radiation era, it is not
slow-roll suppressed in either Poisson or uniform-density gauge.
This becomes the dominant term in the tensor perturbations on large
scales and corresponds to the production of gravitational waves from
density perturbations in the radiation dominated era. Ananda \emph{et al.}
\cite{Ananda:2006af} showed that the amplitude of tensor
perturbations reaches a maximum at Hubble-crossing, $k\eta\approx1$
(where our large-scale approximation breaks down), and thereafter
follows the usual behaviour for free gravitational waves, $|h|
\propto a^{-1}$, as the scalar metric perturbations oscillate and
decay on sub-Hubble scales, $k\eta\gg1$. The first-order scalar
perturbation, $\Phi$, decays rapidly on sub-Hubble scales leaving an
unambiguous prediction for the amplitude of gravitational waves.

Thus we conclude that the power spectrum of the stochastic
gravitational wave background generated from scalar metric
perturbations from inflation is due primarily to the production of
tensor perturbations around Hubble-entry during the radiation era,
and the second-order gravitational waves from inflation are
suppressed in all models of slow-roll inflation.

\begin{acknowledgments}
FA is supported by the Japanese Society for the Promotion of Science
(JSPS). HA is supported by the British Council (ORSAS). KK is
supported by the European Research Council. DW is supported by STFC.
We are grateful to Marco Bruni for useful discussions.
\end{acknowledgments}

\appendix

\section{\label{InverseMetric}Inverse metric up to second order}
From $g_{\mu\lambda}g^{\lambda\nu}=\delta_\mu^\nu$ we can obtain
the inverse metric order by order. At zeroth order we get
\begin{equation}
g^{00(0)}=-a^{-2}, \quad g^{0i(0)}=0, \quad
g^{ij(0)}=a^{-2}\delta^{ij}.
\end{equation}
At first order, it gives
\begin{equation}
g^{00(1)}=2a^{-2}\A^{(1)}, \quad g^{0i(1)}=a^{-2}B^{i(1)},
\quad
g^{ij(1)}=a^{-2}\left(2\C^{(1)}\delta^{ij}-C^{ij(1)}\right).
\end{equation}
The second order components of the inverse metric are
\begin{eqnarray}
g^{00(2)}&=&a^{-2}\left[B_i^{(1)}B^{i(1)}-4\left(\A^{(1)}\right)^2+\A^{(2)}\right],
\\
g^{0i(2)}&=&a^{-2}\left[2\left(\C^{(1)}-\A^{(1)}\right)B^{i(1)}
-B_j^{(1)}C^{ij(1)}+\frac{B^{i(2)}}{2} \right],
\\
g^{ij(2)}&=&-a^{-2}\left[B^{i(1)}B^{j(1)}-4\left(\C^{(1)}\right)^2\delta^{ij}
+4\C^{(1)}C^{ij(1)}-C_k^{i(1)}C^{kj(1)}-\C^{(2)}\delta^{ij}+\frac{1}{2}C^{ij(2)}
\right],
\end{eqnarray}
where the indices were raised with $\delta^{ij}$.


\end{document}